\documentclass[english]{nature}
\makeatletter

\usepackage{amsmath,amsfonts,amssymb}
\usepackage{graphicx}

\begin{document}
\title{\noindent Efficient low-power terahertz generation via on-chip
triply-resonant nonlinear frequency mixing} 
\author{J. Bravo-Abad$^1$, A. W. Rodriguez$^1$, J. D. Joannopoulos$^1$, P. T. Rakich$^2$, S. G. Johnson$^3$, and \\ M. Solja\v{c}i\'{c}$^{1}$}
\maketitle

\vspace{-0.5cm}

\begin{affiliations}
\item Department of Physics, Massachusetts Institute of Technology, Cambridge, MA 02139, USA
\item Sandia National Laboratories, Albuquerque, NM 87185, USA
\item Department of Mathematics, Massachusetts Institute of Technology, Cambridge, MA 02139, USA
\end{affiliations}

\begin{abstract}
Achieving efficient terahertz (THz) generation using compact turn-key sources operating
at room temperature and modest power levels represents one of the critical challeges 
that must be overcome to realize truly practical applications based on THz\cite{ReviewTHzNatMat,Mittlemanbook,ReviewTHzNatPhot}. 
Up to now, the most efficient approaches to THz generation at room temperature -- relying mainly on optical rectification schemes -- require intricate phase-matching set-ups and powerful lasers\cite{Vodopyanov06,Nelson}. Here we show how the unique light-confining properties of triply-resonant photonic resonators can be tailored to enable dramatic enhancements of the conversion efficiency of THz generation via nonlinear frequency down-conversion processes. We predict that this approach can be used to reduce up to three orders of magnitude 
the pump powers required to reach quantum-limited conversion efficiency of THz 
generation in nonlinear optical material systems. Furthermore, we propose a realistic design readily accesible experimentally, both for fabrication and demonstration of optimal THz conversion efficiency at sub-W power levels.
\end{abstract}

The possibility of enhancing the power radiated by an electromagnetic current source by embedding it in a resonant cavity featuring both a high quality factor and small modal volume, has been recognized since the early days of modern electromagnetic theory\cite{Condon41,Purcellbook} as an attractive approach to extract efficiently the energy emitted by different classes of electromagnetic sources. With the emergence of nonlinear optics, this approach was also adopted as a way to enhance a broad variety of nonlinear optical phenomena\cite{Askin66,Boydbook}. Later on, the advent of concepts such as photonic crystals (PhC)\cite{Yablonovitch,SJohn,JJ}, along with the rapid development of improved micro- and nano-fabrication techniques, enabled the demonstration of wavelength-scale counterparts of traditional nonlinear optical devices, which offered the possibility of realistic on-chip active control of light with light itself\cite{ReviewMarin,nonlinearPhCbook}. One of the most remarkable illustrations of this concept resides in on-chip enhancement of the conversion efficiency of nonlinear frequency mixing processes whose conversion efficiency is intrinsically very low. Of special interest, due to its considerable importance for applications, is the case of efficient terahertz (THz) generation. Recently, different approaches to resonant enhancement of difference-frequency nonlinear coupling processes, including those involving a final frequency in the THz regime, have been proposed\cite{Avetisyan99,Morozov05,Ianlong}. In this paper, we present a scheme that enables enhancement
of THz power generation via second-order nonlinear frequency
downconversion by up to three orders of magnitude compared to conventional nonresonant approaches. By using a combination of accurate numerical simulations and a rigorous coupled-mode theory, we show how the unique properties of photonic microresonators to confine light in subwavelength
volumes for many optical periods can be tailored to dramatically reduce the pump power
at which the theoretical maximum of THz generation predicted by the Manley-Rowe quantum limit\cite{Boydbook} is reached. Furthermore, we show how a generalization of the canonical phase-matching conditions enables efficient THz generation in nonlinear optical
materials in which conventional phase-matching techniques are difficult, or even impossible, to implement.
Finally, we propose an implementation of
highly-efficient compact on-chip THz sources operating at room temperature and pumped by sub-$\mu$J energy pulses, thus opening
the way towards practical realization of a number of important applications in a broad 
spectrum of fields, ranging from communications to biomedical imaging\cite{ReviewTHzNatMat,Mittlemanbook,ReviewTHzNatPhot,Everitt}.

To gain physical insight into the approach introduced in this manuscript, we start by revisiting the enhancement of the power radiated by
an arbitrary current source ${\bf J}_T({\bf r},t)$ embedded in a single-mode lossless dielectric cavity. We assume that the electric field inside the cavity can be written as  ${\bf E}({\bf r},t)=a_T(t)\:\exp(- i \omega_T t)\:{\bf E}^{(0)}_T({\bf r})/\sqrt{U}$ (where $E^{(0)}_T({\bf r})$ is the electric field profile of the resonator mode, $U=\int d{\bf r} \: \epsilon_0\: n_T^2({\bf r}) |{\bf E}_T^{(0)}({\bf r})|^2$ is the electromagnetic energy stored in the resonant mode, and $a_T(t)$ is the corresponding slowly-varying electric field amplitude\cite{Haus}). If we also assume that ${\bf J}_T({\bf r},t)={\bf J}_T^s({\bf r},t)\:\exp(-i \omega_T t) $ and that all the radiated power is collected by means of a waveguide coupled evanescently to the cavity, the emitted THz power ($P_T$) can be written in the undepleted regime as
\begin{equation}
P_T  =4\left( \frac{n_T}{\pi c_0\epsilon_0 \lambda_T^2} \right)
\left( \frac{Q_T}{Q_{T,s}} \right)
\left( \frac{Q_T}{\widetilde{V}_T} \right)
\left| \int d{\bf r}\:{\bf J}_T^s({\bf r},t) \: [{\bf E}^{(0)}_T({\bf r})]^*/E_{T,\textrm{max}}  \right|^2
\end{equation}
where $Q_T$ and $Q_{T,s}$ stand for the total quality factor and the external quality factor (i.e., the one governing the decay into the waveguide) of the resonator, respectively. $\widetilde{V}_T=V_T/(\lambda_T/n_T)^3$, where $V_T$ is the effective modal volume of the resonator; whereas $E_{T,max}$ denotes the maximum value of $|{\bf E}_T({\bf r})|$.  

From Eq. (4), the enhancement of the power radiated by ${\bf J}_T({\bf r},t)$ inside the cavity is apparent through the factor $Q_T/\widetilde{V}_T$. Note that a $Q_T/\widetilde{V}_T$ enhancement factor for sources in resonant cavities is famous from Purcell enhancement of spontaneous emission\cite{Purcell46}, but also appears in broad range of purely classical contexts, from microwave antennas inside cavities to acoustic effects such as a blown flute or bowed violin. As we discuss in detail below, it is precisely this enhancement factor, together with 
the unprecedented large values for $Q_T/\widetilde{V}_T$ that can be realized in photonic microresonators\cite{Noda,Vahala_Review}, that enable increase of the conversion efficiency of optical nonlinear frequency conversion processes to an extent that can not be achieved by means of any other currently known physical mechanism. Note also that Eq. (4) reflects the fact that in order to maximize the output power $P_T$, the intrinsic absorption rate inside the cavity must be matched to the decay rate to the waveguide (i.e., for a given value of the $Q$-factor characterizing the intrinsic absorption losses inside the cavity, $Q_{abs}$, the product $Q_T/Q_{T,s}\times Q_T$ appearing in the right-hand side of Eq. (4) is maximized when $Q_{T,s}=Q_{abs}$). This $Q$-matching condition for the efficient extraction of the power generated inside the cavity is a general property that appears in a broad range of different contexts (thermal, electromagnetic, mechanical, etc) that require efficient emission of energy from a high-$Q$ system\cite{JJ,Haus,Qmatchingref}. 

In the case of a general nonlinear frequency mixing process involving a pump and idler frequencies (denoted hereafter by $\omega_1$ and $\omega_2$, respectively) in a $\chi^{(2)}$ crystal, the current distribution ${\bf J}_T({\bf r},t)$ arises from the temporal variation of the nonlinear polarization vector ${\bf J}_T({\bf r},t)= \partial {\bf P}^{NL}({\bf r},t)/ \partial t$. In particular, in the case of  difference-frequency generation (DFG), if we assume that the electric fields at $\omega_1$ and $\omega_2$ are given by ${\bf E}_1({\bf r},t)=a_1(t)\:\exp(- i \omega_1 t)\:{\bf E}^{(0)}_1({\bf r})/\sqrt{U_1}$ and ${\bf E}_2({\bf r},t)=a_2(t)\:\exp(- i \omega_2 t)\:{\bf E}^{(0)}_2({\bf r})/\sqrt{U_2}$,  respectively (where $U_i$ is the electromagnetic energy stored in the system at frequency $\omega_i$), the expression for the power $P_T$ emitted at the final frequency $\omega_T=\omega_1-\omega_2$ takes the following form
\begin{equation}
P_T=\frac{2^4 c_0 \pi n_T}{\epsilon_0 \:\lambda_T^4}\frac{Q_T}{Q_{T,s}}\frac{Q_T}{\widetilde{V_T}}|a_1(t)|^2\:|a_2(t)|^2 |\beta_{\textrm{eff}}|^2
\end{equation}
where $\beta_{\textrm{eff}}$ represents the nonlinear coupling strength between the electromagnetic fields involved in the nonlinear difference-frequency mixing. This magnitude can be written as
\begin{equation}
\beta_{\textrm{eff}}=\frac{\int{d{\bf r}\: \sum_{i=1,2,3} [E_{Ti}^{(0)}({\bf r})]^* / E_{T,\textrm{max}}} \: \left\{\sum_{j,k=1,2,3} 
\chi_{ijk}^{(2)}({\bf r})  E_{1j}^{(0)}({\bf r})\: [E_{2k}^{(0)}({\bf r})]^* \right\}} 
{\sqrt{\int{d{\bf r} \: n_1^2({\bf r})\:|{\bf E}^{(0)}_1({\bf r})|^2}} \: \sqrt{\int{d{\bf r} \: n_2^2({\bf r})\:|{\bf E}^{(0)}_2({\bf r})|^2})}}
\end{equation}
where $\chi_{ijk}^{(2)}({\bf r})$ stands for the spatial distribution of the second-order nonlinear susceptibility tensor (here the
subindices $\{i,j,k\}$ stand for the components $\{x,y,z\}$, respectively, of the corresponding electric field vectors).

An important figure of merit for the scheme proposed in this manuscript is the enhancement factor ($\eta_{\textrm{enh}}$) of the output power $P_T$ predicted by Eq. (5) with respect to value of $P_T$ that one would obtain using traditional nonresonant approaches to enhance the conversion efficiency of a DFG process. In particular, we find relevant the comparison of our scheme with the case in which the nonlinear coupling coefficient ($\beta_{\textrm{eff}}$) is increased simply by reducing the nonlinear interaction area of the modes involved in the frequency mixing process, using for instance a waveguide for the pump, idler and final frequencies (an approach that has been used extensively in the past for enhancing the conversion efficiency of DFG in different wavelength regimes\cite{Askin66,Boydbook}). For simplicity, in this comparison we assume operation in the undepleted regime (i.e., we take $a_1(t)=a_1(0)$ and $a_2(t)=a_2(0)$ in Eq. (5)). Additionally, we assume that in our system both the pump and idler fields are temporally confined, this confinement being characterized by $Q$-factors $Q_1$ and $Q_2$ for $\omega_1$ and $\omega_2$, respectively. As we discuss below for a particular potential implementation of our scheme, this temporal confinement not only permits enhancing $P_T$, but also allows introducing a generalization of the canonical phase-matching condition\cite{Boydbook}, which leads to efficient DFG process even in systems in which the implementation of standard phase matching techniques (such as birefringence) is difficult or even impossible. Keeping these assumptions in mind, after some algebra, one finds that it is possible to write an accurate analytical approximation for $\eta_{\textrm{enh}}$ as
\begin{equation}
\eta_{\textrm{enh}}\approx \frac{2^5 \lambda_1 \lambda_2}{\pi^3 \lambda_T^2}\: \frac{n_Tn_{T,\textrm{eff}}\:n_{1,\textrm{eff}}\:n_{2,\textrm{eff}}}{n_1^2 n_2^2}
\: \frac{A_{\textrm{wg}}}{L_{\textrm{wg}}^2}\: \frac{Q_T}{Q_{T,s}} \:\frac{Q_T}{\widetilde{V}_T}\:Q_1 Q_2
\end{equation}
where $\lambda_1$, $\lambda_2$, and $\lambda_T$ stand for the wavelengths corresponding to the pump, idler, and final frequencies, respectively. $n_1$, $n_2$, and $n_T$ denote the value of the refractive indexes at $\omega_1$, $\omega_2$, and $\omega_T$, respectively, of the dielectric medium where the frequency mixing takes place. The parameters $n_{1,\textrm{eff}}$, $n_{2,\textrm{eff}}$, and $n_{T,\textrm{eff}}$ represent the effective refractive indexes of the waveguide configuration for the corresponding modes at $\omega_1$, $\omega_2$, and $\omega_T$, respectively. $A_{\textrm{wg}}$ is the
transversal area of waveguide system, while $L_{\textrm{wg}}$ stands for the corresponding length of 
the waveguide. Note also that, in order to make a meaningful comparison between the cavity system and its waveguide counterpart, when deriving Eq. (7), we have assumed that in both cases there exists an optimal overlap (which fulfills the phase-matching condition for each case) between ${\bf E}_1({\bf r},t)$, ${\bf E}_2({\bf r},t)$, and ${\bf E}_T({\bf r},t)$.

Equation (7) summarizes well the comparison between the different mechanisms that play a role to enhance the conversion efficiency of a DFG process in a waveguide configuration (considering the optimal case in which phase-matching condition between the waveguide modes is satisfied) and the scheme to enhance the conversion efficiency introduced in this manuscript. In particular, we have found that, as we numerically demonstrate below for a particular structure, in the case of THz generation via a $\chi^{(2)}$ DFG process in realistic nonlinear optical material configurations (in which considering the absorption losses at the final THz frequency is a key aspect in determining the ultimate conversion efficiency), it is possible to reach values $\eta_{\textrm{enh}} \sim 10^3$. We emphasize that this large value for $\eta_{\textrm{enh}}$ is obtained even if the corresponding phase-matching condition is satisfied in the waveguide system, which is often challenging to implement due to the vast difference between the wavelengths corresponding to the pump and idler modes and the final THz modes. Thus, we believe that the scheme analyzed in this work could be of paramount importance for increasing the conversion efficiency of nonlinear frequency conversion processes whose conversion efficiency is intrinsically low due, for instance, to the lack of a phase-matching mechanism in the considered frequency range, or due to the small value of $\chi^{(2)}$ of the materials of interest. This conclusion is the first important result of this manuscript.

In order to explore the extent to which this concept can be applied to solve the current lack of efficient sources and detectors
operating at room temperature in the so-called THz frequency gap\cite{ReviewTHzNatMat,ReviewTHzNatPhot}, we illustrate its implementation in a specific structure based on a triply-resonant nonlinear configuration. Figure 1a displays a schematic of the proposed system. The power carried by two NIR beams of wavelengths $\lambda_1$ and $\lambda_2$ (playing the role of idler and pump beams, respectively, their corresponding power being $P_{1in}$ and $P_{2in}$) is coupled, by means of an index-guided waveguide, to two high-order whispering gallery modes (WGM) supported by a dielectric ring resonator. These WGM at $\lambda_1$ and $\lambda_2$ are characterized by angular momenta $m_1$ and $m_2$, respectively. The ring resonator also acts as a dipole-like defect at $\lambda_T$, when embedded in an otherwise perfectly periodic THz-wavelength scale photonic crystal (PhC) formed by a square lattice of dielectric rods (see the corresponding electric field profile in Fig. 1b). Thus, the $\chi^{(2)}$ nonlinear frequency down-conversion interaction that takes place between the two NIR WGM's circulating inside the ring resonator yields a current distribution that radiates inside the PhC cavity at the frequency difference $\omega_T=\omega_1-\omega_2$; the rate at which
the radiation is emitted is strongly enhanced by the PhC environment in which the ring resonator is embedded. In order to extract efficiently the THz output power ($P_T$) from the PhC cavity, we introduce into the system a PhC waveguide created by reducing the radius of a row of rods (see Figs. 1a and 1b). In addition, in order to break the degeneracy existing between the $x$-and $y$-oriented dipole defect modes, the radius of two of the nearest neighbors rods of the ring resonator is reduced with respect to the radius of other rods in the PhC. This configuration permits having a large value for factor ($Q_T/\widetilde{V}_T$), along with a high-$Q$ resonant confinement also for the pump and idler frequencies.

Figure 1b shows the structure that results from optimizing the geometrical parameters of the system for efficient generation at 1 THz, assuming a pump beam of wavelength $\lambda_1$=1550nm, an idler beam with $\lambda_2$=1542nm,  and that the structure is implemented in GaAs (in which the relevant component of the nonlinear susceptibility tensor is $d_{14}$=274pm/V). Note that because the resonant cavity in our design forms a filter for the infrared (IR) input power, extremely tight control of closely spaced source-frequency lines is not required to produce a specific THz frequency, as long as the input spectrum overlaps with the resonant filter bandwidth (although greater frequency control and hence greater overlap leads to higher efficiency). Importantly, we also point out that in order to maximize the strength of the nonlinear coupling coefficient that governs the energy transfer between the pump, idler and THz fields, the whole structure must be designed so the dependence of the THz electric field profile on the azimuthal coordinate $\theta$ inside the ring resonator (see definition of $\theta$ in inset of Fig. 1b) cancels the modulation introduced in the nonlinear susceptibility tensor by the local variation of the pump and idler fields with respect to the axes of the nonlinear crystal \cite{OLSipe,Dumeige}. This modulation is given by the dependence on $\theta$ of the product ${\bf E}^{(0)}_1({\bf r})\:{\bf E}_2^{(0)*}({\bf r})$, which in the case of two considered WGMs is given by a factor $\exp\left[\imath\theta(m_2-m_1\pm 2)\right]$\cite{footnote4}(see inset of Fig. 1c). For GaAs, and for the above cited values for $\lambda_1$ and $\lambda_2$, we have found that this condition is fulfilled by two WGM with $m_1$=572 and $m_2$=575, and a dipole defect mode in the THz-scale PhC. 

The above discussion can be viewed as a generalization of the canonical phase-matching condition often found in nonlinear optics\cite{Boydbook}. Specifically, note that in standard phase-matching techniques, the dispersion relation for the three frequencies involved in the DFG is exclusively determined by the intrinsic properties of the particular material in which the nonlinear interaction takes places. Consequently, in these traditional schemes, the overall efficiency of a DFG process relies entirely on finding a suitable nonlinear material whose dispersion relation permits fulfilling simultaneously, for the frequency range of interest, both the frequency-matching and the phase-matching conditions (or alternatively, on finding some physical mechanism,
such as quasiphase-matching, that permits to phase-match the different fields involved in the nonlinear process). However, in the approach introduced here, the dispersion relation corresponding to the final frequency $\omega_T$ is different from that corresponding to $\omega_1$ and $\omega_2$ and, importantly, it can be tailored almost at will simply by modifying the geometrical parameters that define the THz-scale PhC. This introduces a general and versatile route to phase-matching that does not depend exclusively on the intrinsic properties of naturally existing nonlinear optical materials. This fact could be particularly relevant in those systems in which the canonical phase-matching condition can not be fulfilled.

To compute accurately the nonlinear optical dynamics of the structure sketched in Fig. 1a, we have applied a temporal coupled-mode theory (TCMT) formalism similar to that described in Refs. ~\citenum{Alex,Rafif,Hila,Ianlong}. It has been extensively shown\cite{Alex,Rafif,Hila,Ianlong} that this theoretical framework permits characterizing accurately several nonlinear frequency mixing processes, including those in which there exists a large difference between the wavelength corresponding to the pump and the final frequency\cite{footnote5}. Figure 2 summarizes the results obtained in the continuous-wave (cw) regime. In these calculations we have assumed that $P_{1in}=P_{2in}$ (the dependence of the results on the ratio $P_{2in}/P_{1in}$ is discussed below) and quality factors $Q_1=Q_2=3.5\times10^5$ and $Q_T=10^3$. These values for $Q$ are compatible with both the absorption coefficient of GaAs at 1 THz ($\alpha$=0.5cm$^{-1}$)\cite{Hebling} and the experimental values for the quality factor obtained in similar configurations for the considered ring resonator and the photonic crystal cavity\cite{Noda,Lipson}. As shown in Fig. 2a, for values of 
$P_{1in} >0.01P_0$ (see \cite{footnote7}) the conversion efficiency (defined here as ratio between the output power at THz 
and total input power at NIR frequencies) starts departing from the conversion efficiency predicted by the undepleted approximation, eventually
reaching the maximum value predicted by the Manley-Rowe relation \cite{Boydbook}. Specifically, from the steady-state solution of the TCMT equations, one can find that the maximum conversion efficiency can be written as $\eta_{max}=\left[1/(1+\omega_1^2/\omega_2^2)\right](\omega_1/\omega_T) \Gamma_T \Gamma_1$, where $\Gamma_T=Q_T/Q_{T,s}$ and $\Gamma_1=Q_1/Q_{1,s}$ (see horizontal gray line in Fig. 2a). This fact is confirmed by our numerical simulations (see Fig. 2a). As clearly shown in Fig. 2a, at the critical value of $P_{1in}$ at which this maximum conversion efficiency is reached (in our case $P_{1in}^c=0.19P_0$, or equivalently  $P_{1in}^c=$0.32W) the pump power that is coupled to the ring resonator is completely down-converted inside the system to power at THz and idler frequencies, giving rise to a sharp minimum in $P_{1tr}$ and a maximum in $P_{2tr}$. In the general case in which $P_{1in} \neq P_{2in}$, we have found that the critical powers at which
the maximum conversion efficiency takes places (denoted as $P_{1in}^c$ and $P_{2in}^c$), satisfy the following expression
\begin{equation}
\frac{P_{2in}^c}{P_0}=\frac{\omega_2}{4 \omega_1} \frac{\Gamma_1}{\Gamma_2} \left|1-\frac{P_{1in}^c}{P_0}\right|^2
\label{eq:criticalpower}
\end{equation}
This function is plotted as a solid red line in Fig. 2b. As clearly demonstrated in Figs. 2a and 2b (see vertical dashed line in both
figures), the maximum conversion efficiency occurs at the intersection of $P_{2in}=P_{1in}$ with the expression given in Eq. (6). Note also that that as $P_0 \propto 1/Q_1 Q_2 Q_T$, one can adjust the value of $P_{1in}^c$ just by varying the product $Q_1 Q_2 Q_T$. On the other hand, applying an analysis similar to the one used in Ref.~\citenum{Ianlong}, we have also found that there exists a large region of the parameter space $\{ P_{2in}, P_{1in} \}$ in which the considered structure presents multistable response, i.e., where there are more than one steady-state solution for a given value of the ratio $P_{2in}/P_{1in}$ (see blue area in Fig. 2b). We also point out that the net effect of the absorption losses in the conversion efficiency consists simply in downscaling the results obtained in the lossless case by a factor $\Gamma_T \Gamma_1$ (see dotted line in Fig. 2b).

In order to completely characterize the THz generation process in the analyzed structure we have also studied the temporal evolution of the response of the system to Gaussian pulse excitations. In these calculations we assume that the temporal width of the pulses corresponds to the lifetime of the THz-scale cavity ($\large{\tau}_{\small{THz}} \approx 16$ns). The value of $\large{\tau}_{\small{THz}}$ is much larger than the lifetime of the WGM modes at the pump and idler frequencies ($\approx 0.8$ns). Thus, we expect similar conversion maximum effiencies as those found in the above cw analysis. Figures 3a-c show the results corresponding to three representative values for the peak power of $P_{1in}(t)$ (labeled as A,B and C, respectively, in Fig. 2b). As seen in Fig. 3a, when the maximum $P_{1in}(t)$ is equal to the critical power $P_{1in}^c$, the pump input pulse is completely consumed after spending approximately 60ns in the system (i.e., $P_{1tr} \approx 0$ after 60ns); and, simultaneously, the power of the transmitted idler pulse ($P_{2tr}(t)$) value reaches a peak value that is approximately twice the peak value corresponding to the input idler pulse ($P_{2in}(t)$). For peak values of $P_{1in}(t)$ much lower than the critical power $P_{1i}^c$, the almost undepleted behavior can be clearly observed (see Fig. 3b): the peak powers of the pump and idler pulses pulses are barely modified as they travel through the system. On the other hand, Fig. 3c clearly shows how, for input NIR peak powers well beyond the critical power, the pump pulse is completely consumed at $t \approx 40$ns (see inset of Fig. 3c). However, in contrast to the case displayed in Fig. 3a, Fig. 3b shows how after $t \approx 40$ns the up-conversion process that mix $\omega_T$ with $\omega_2$ to yield  $\omega_1$  starts being relevant, and, consequently, $P_{1tr}(t)$ begins increasing with $t$; which in turn reduces the overall THz conversion efficiency of the process. Finally, Fig. 3d displays a summary of our time-dependent simulations in terms of the ratio between the output THz energy and total input NIR energy (defined as 
$E_{THz}=\int_{-\infty}^{\infty} dt\:P_T(t)$, and $
E_{NIR}=\int_{-\infty}^{\infty} dt\:\left[P_{1in}(t)+P_{2in}(t)\right] $, respectively). As displayed in Fig. 3a, the maximum conversion efficiency can be reached for an input energy $E_{NIR}=$0.04$\mu$J, which
represents a reduction in $E_{NIR}$ of three orders of magnitude with respect to the most efficient schemes for THz generation 
in nonlinear crystals reported up to date \cite{Nelson}. Furthermore, note that, in addition to powerful lasers, current efficient schemes for THz generation require intricate phase-matching set-ups, whereas in the system introduced in this manuscript maximum
theoretically possible efficiency can be achieved in an integrated structure having a total area of approximately 1mm$^2$.

In conclusion, we have shown the dramatic enhancement of the conversion efficiency of general difference-frequency downconversion processes enabled by triply-resonant photonic resonators. In addition, we have presented a generalization of the canonical phase-matching condition, which makes possible efficient DFG processes even in the case of nonlinear optical materials where standard phase-matching techniques are difficult or impossible to implement in certain frequency regimes. By means of detailed numerical simulations, we have illustrated the relevance of the proposed scheme by demostrating complete conversion to THz energy of a 0.04$\mu$J NIR pump pulse in a realistic 1mm$^2$-footprint on-chip structure created from GaAs. Alternatively, we have demonstrated that in the continuous-wave regime the pump powers required to reach quantum-limited conversion efficiency can be reduced to up three orders of magnitude with respect to the conventional approaches for THz generation employed up to date. In contrast to previous high-efficiency THz generation schemes, the concept introduced in this manuscript opens, for the first time, the way to efficient THz generation from sources that are compact, turn-key, and low-cost, which we believe could enable a broader use of THz sources.

The authors thank Dr. Morris Kesler and Dr. Katie Hall for valuable discussions. This work was supported by the MRSEC
Program of the National Science Foundation under award number DMR-0819762 and by the U.S. Army Research Office
through the Institute for Soldier Nanotechnologies under Contract No. W911NF-07-D-0004.

\begin{figure}[htb]
	\centering
\includegraphics[width=8.7cm]{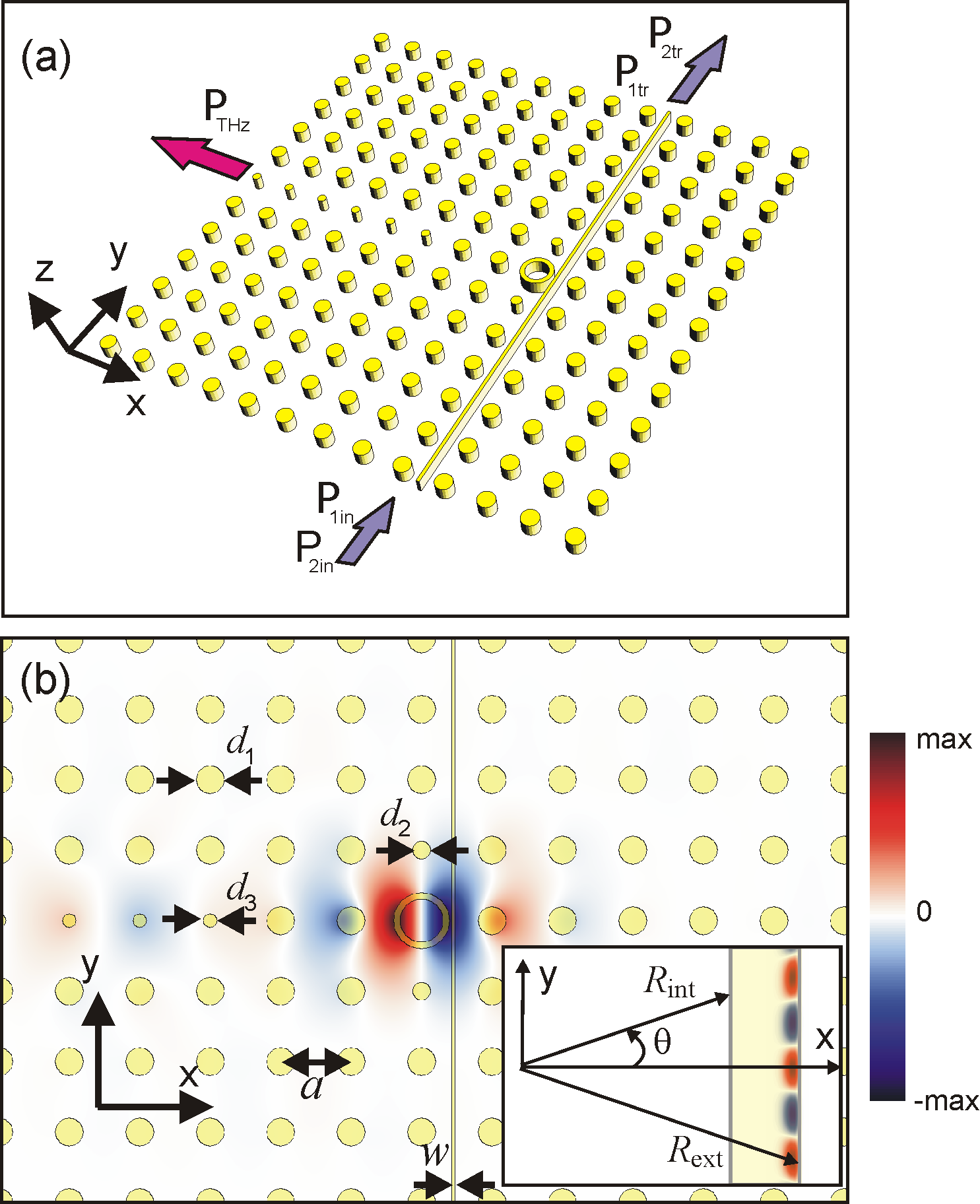}
\caption{(a) Schematic of the triply-resonant nonlinear photonic structure analyzed in the text. $P_{1in}$ and $P_{2in}$ denote the
input powers at the pump and idler frequencies, respectively; whereas  $P_{1tr}$ and $P_{2tr}$ represent the corresponding transmitted powers through the structure. $P_{THz}$ stands for the THz ouput power. (b) Electric field profile ($E_z$) as obtained from FDTD calculations corresponding to the resonant mode appearing at 1THz in the structure shown in Fig. 1a. The value of the different geometrical parameters displayed in this figure are $a$=102$\mu$m, $d_1$=40.8$\mu$m, $d_2$=5.3$\mu$m, $d_3$=4.0$\mu$m, and $w$=0.8$\mu$m. Inset displays an enlarged view of the electric field profile (pointing along the $z$-direction) corresponding to a whispering gallery with $m$=572 circulating inside the dielectric ring shown in the main figure. The geometrical parameters defining the ring resonator are also shown in the inset, $R_{ext}$ and $R_{int}$
being 40.1$\mu$m and 30.5$\mu$m, respectively. Yellow areas in both the main and inset figures represent GaAs regions, while
white areas represent air.}
\label{f:schematic}
\end{figure}

\begin{figure}[htb]
	\centering
\includegraphics[width=8.7cm]{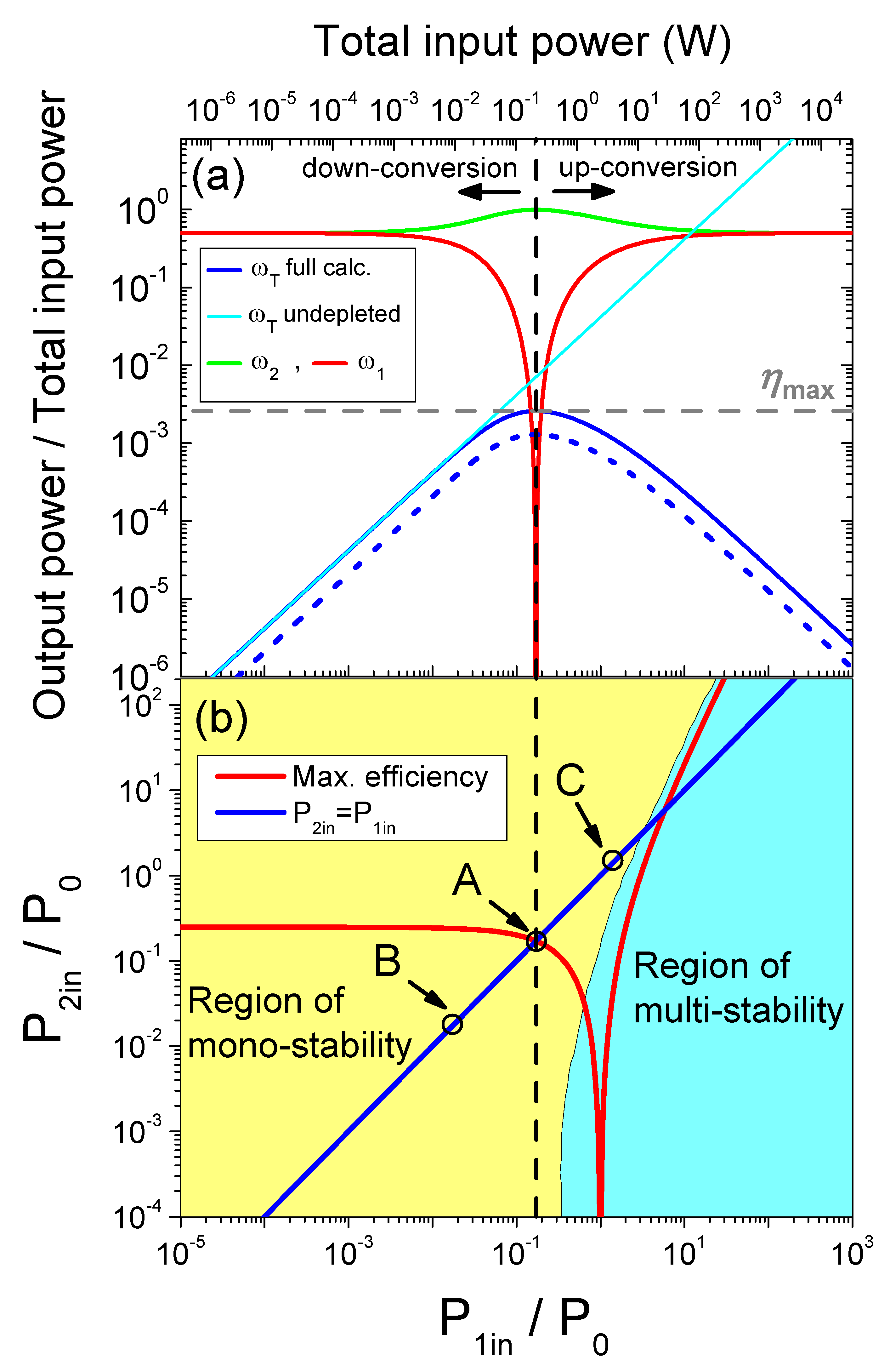}
\caption{(a) Ratio between the total output power emitted by the system at 1 THz and the total input power at the NIR pump and idler frequencies.
The results for the three frequencies involved in the considered nonlinear down-conversion process are displayed ($\omega_1$, $\omega_2$, and $\omega_T$ correspond to the pump, idler, and final THz frequencies, respectively). (b) Solid line renders the dependence between $P_{2in}$
and $P_{1in}$ that yields the maximum THz conversion efficiency in the analyzed configuration. Blue line represents the case $P_{2in}=P_{1in}$.
Yellow and blue areas represent the regions of mono-stability and multi-stability, respectively, in the space of parameters $\{P_{1in},P_{2in}\}$. Labels A,B, and C correspond to the time-dependent analysis displayed in Figs. 3a, 3b, and 3c, respectively.}
\label{f:schematic}
\end{figure}

\begin{figure}[htb]
	\centering
\includegraphics[width=14cm]{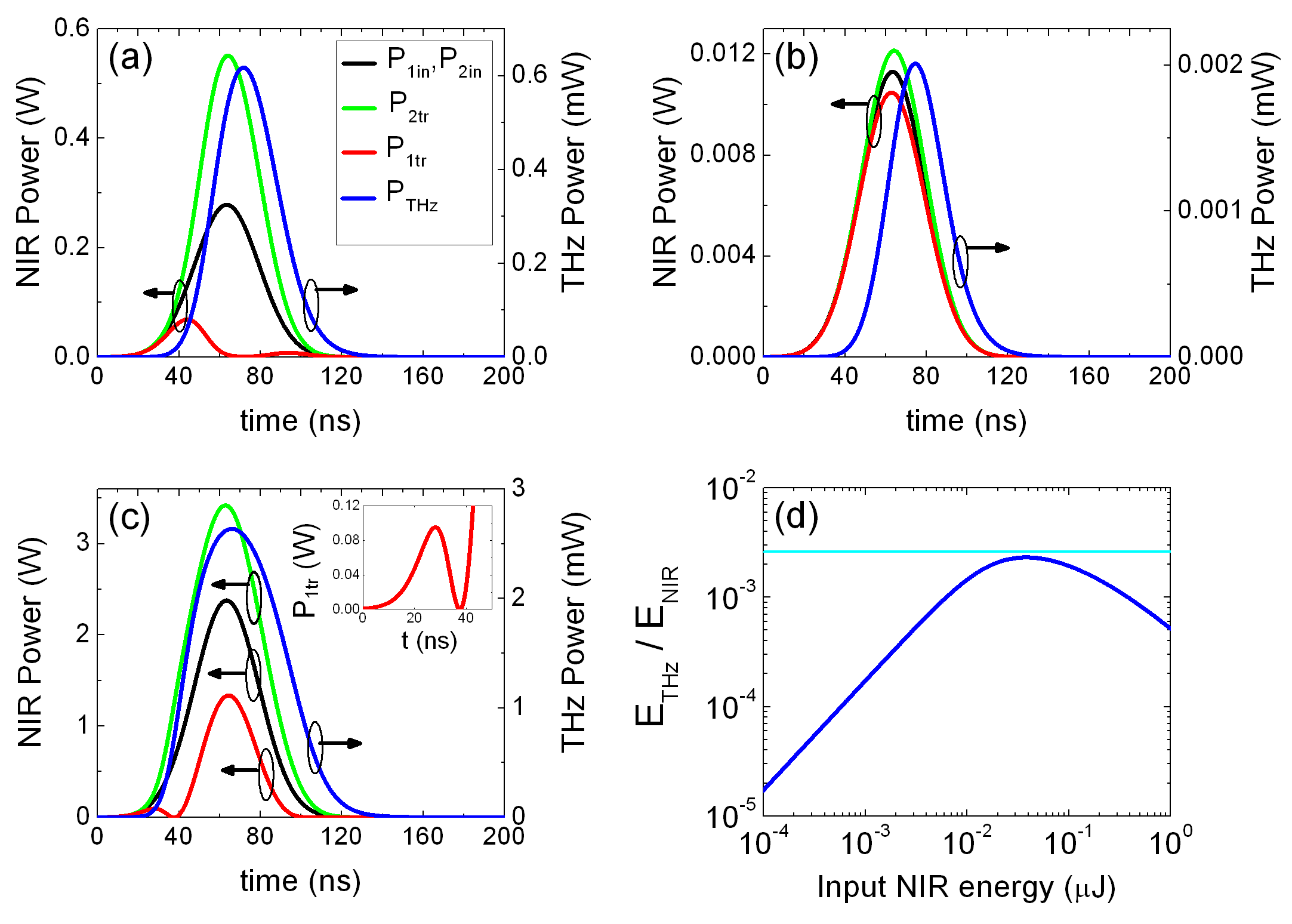}
\caption{Analysis of the temporal response of the system shown in Fig. 1a to gaussian excitation pulses. Panels a, b, and c correspond to 
the peak values for $P_{1in}$ shown by labels A,B, and C, respectively in Fig. 2b. In Panels a, b, and c, the scale on the left vertical axis corresponds to the NIR power of the pump and idler frequencies, both for the input and transmitted pulses; whereas the scale on the right right vertical axis corresponds to the THz output power. Inset of Fig. 3c, shows an enlarged view of the temporal dependence of $P_{1tr}(t)$ between
$t=0$ and $t=50$ns. Panel d displays the ratio between the total output energy ($E_{THz}$) and the total NIR input energy ($E_{NIR}$) as a function of $E_{NIR}$. Horizontal line in the inset displays the maximum possible conversion efficiency
given by the Manley-Rowe quantum limit.}
\label{f:schematic}
\end{figure}


\begin{thebibliography}{}
\bibitem{footnote3b}
This value for the THz frequency was chosen for illustration purposes, but the approach proposed in this manuscript is general and we have obtained similar results as those shown in this manuscript for different final THz frequencies.
\bibitem{footnote5}
Standard numerical methods used in nonlinear nanophotonics (such as the nonlinear FDTD) are not suitable for being
applied to this problem, mainly due to the vast difference between the wavelength corresponding to the pump and final THz frequency (the THz wavelength is about 2 orders of magnitude larger than the pump wavelength in the cases of interest).
\bibitem{footnote7}
Following Ref.~\citenum{Ianlong}, we have defined $P_0=\omega_1 \Gamma_1/(4 |\beta|^2 Q_1 Q_2 Q_{THz})$, where $\beta$ stands for the nonlinear coupling coefficient. For the particular structure analyzed in the text $\beta=1.2\sqrt{J}$ or equivalently $P_0=1.6$W. Note also that $P_0$ corresponds to the critical power in which the maximum conversion efficiency is reached when $P_{2in} \to 0$.
\end{thebibliography}
\end{document}